# Internal structure of nanoparticles of Al generated by laser ablation in liquid ethanol

G. Viau,[a] V. Collière,[b] G.A. Shafeev[c]

[a] Université de Toulouse, INSA-LPCNO, CNRS UMR 5215, IRSAMC, 135, av. de Rangueil, F-31077 Toulouse Cedex 4, France

[b] Université de Toulouse, CNRS UPR 8241, Laboratoire de Chimie de Coordination, 205 route de Narbonne, F-31077 Toulouse, France

[c] Wave Research Center of A.M. Prokhorov General Physics Institute of the Russian Academy of Sciences, 38, Vavilov Street, 119991 Moscow, Russian Federation

Abstract

Al NPs are synthesized by laser ablation of a bulk Al target immersed into liquid ethanol saturated with $H_2$ at atmospheric pressure. The nanoparticles possess a well-distinguished core-shell structure. High Resolution Transmission Electron Microscopy shows several layers inside the Al nanoparticle: oxide layer, amorphous Al, single crystal Al, and a cavity in the center. Formation of the cavity is attributed to the sharp increase of hydrogen dissolution in Al upon its melting and its eventual release after the solidification.

Recently, laser-assisted nanostructuring of Al by laser ablation in liquids attracted much attention. This technique provides the possibility of generating a large variety of NPs that are free of both surface-active substances and counter-ions (see [1] and references therein). It has been demonstrated that this process leads to the formation of self-organized nanostructures on the bulk Al surface [2-4], while some material of the target is dispersed into the surrounding liquid as NPs of metallic Al [5]. High reactivity of Al with ambient oxygen and water can partially be compensated by ablation using short laser pulses due to faster quenching which could minimize the formation of either oxide or hydroxide on the particle surface. Furthermore, oxidation of NPs generated by this technique can be additionally suppressed by carefully outgassing the air dissolved in liquid and/or replacement of it by a selected gas.

Laser ablation of a bulk Al target in ethanol results in the formation of stable colloid of metallic Al NPs without addition of any surface-active substances [5]. The obtained NPs were found mostly amorphous with several (3-4) crystalline inclusions per particle. Moreover, some Al particles prepared by this method presented an inner cavity. The preliminary hypothesis of the origin of these cavities was the interaction of the molten Al NPs with traces of water in the surrounding ethanol resulting in the formation of hydrogen [5]. Indeed, the gas dissolved in the liquid in which the laser ablation is carried out may alter the properties of generated NPs. In a typical laser ablation experiment the composition of gases dissolved in the liquid is not controlled and, therefore, these gases are close to air in composition. Both $O_2$ and $N_2$ are not highly soluble in metals. On the contrary, the solubility of light gases, e.g., hydrogen, in metals is high, especially in liquid ones. Metallic NPs possess high specific surface and dissolution of the gas in them can be very efficient. Saturation of the liquid in which the laser ablation is carried out with a gas would change the chemical composition of the vapor around the liquid NPs during their ejection from the target and therefore may offer the possibility to control over their properties.

In this communication we report on the morphology of Al NPs obtained by laser ablation of a bulk target in liquid ethanol saturated with hydrogen. HR TEM imaging reveals the internal structure of Al NPs generated by laser ablation of a bulk target in ethanol saturated with $H_2$.

A Nd:YAG laser at the wavelength of 1.06 μm, 10 Hz repetition rate and the pulse width of 30 ps was used for generation of Al NPs in absolute ethanol saturated by gaseous hydrogen. For the ablation in anaerobic conditions the target was placed into a sealed-off Pyrex cell, filled with absolute ethanol and outgassed with a vacuum pump following several cycles of cell freezing with liquid nitrogen. The liquid in the cell was purged with gaseous hydrogen at ambient pressure prior to laser exposure for some time to saturate the liquid. Laser radiation was focused through the Pyrex cap and liquid layer on the target and the typical exposure time was approximately 1 hour at typical fluence on the target of 4 J/cm$^2$. Hydrogen purge was kept on during the whole laser exposure. High Resolution Field Emission Transmission Electron Microscopy (HRTEM) was performed on a Jeol JEM 2100F UHR at 200 kV equipped with a field emission electron source. The optical absorption spectra of the colloidal solution were recorded in the range of 200-800 nm with the help of a Perkin-Elmer spectrophotometer.

The general view obtained with a Transmission Electron Microscope of these NPs generated in liquid ethanol is presented in the Figure 1. The particles produced by laser ablation of an aluminium target in ethanol are spherical as inferred from transmission electron microscopy (Fig. 1). The image contains lighter and darker areas, and their arrangement can be deduced from the images of the same particle taken at different tilts of the microscope grid with respect to the axis of the electron beam.

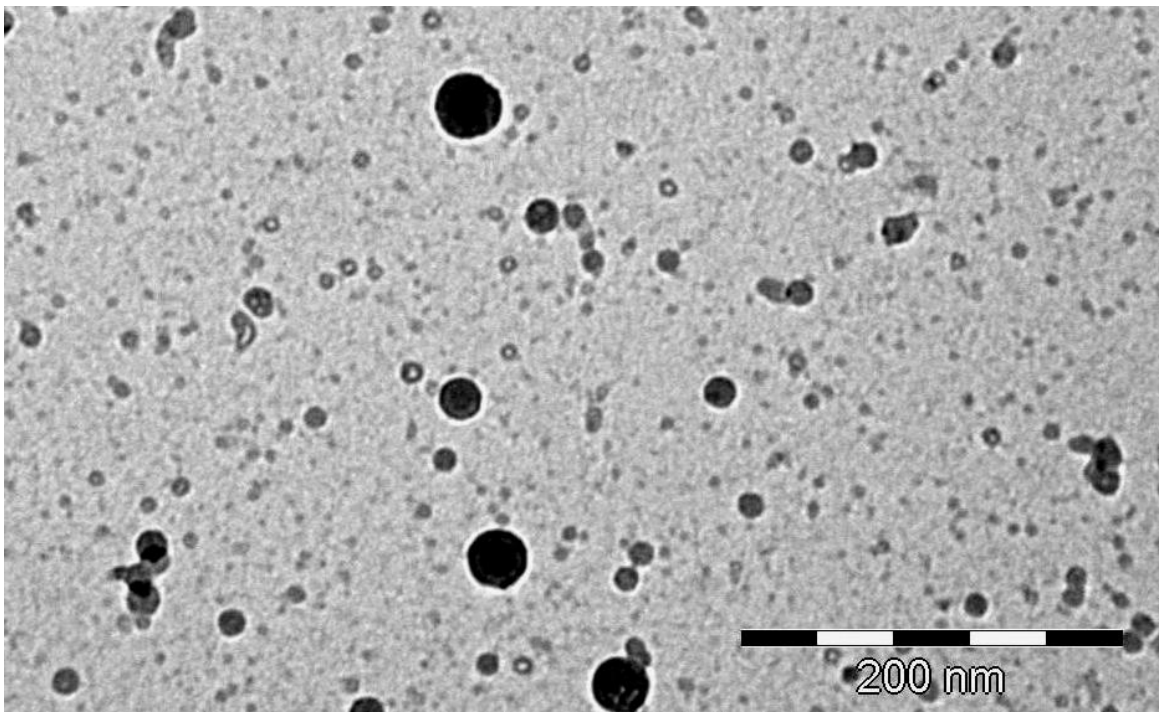

Fig. 1. TEM view of Al NPs generated by ablation of a bulk Al target in liquid ethanol saturated with $H_2$. Scale bar denotes 200 nm.

The average size of particles is 30 nm as inferred from TEM images. Most of these particles present an inner cavity. The bright part within the particle is explained by a lower absorption of the electron beam due to a lack of matter. One can see that the light region of the image remains inside the contour of the particle in a wide range of tilting angles. One may conclude that this light region is situated inside the particle and is not a depression on its surface.

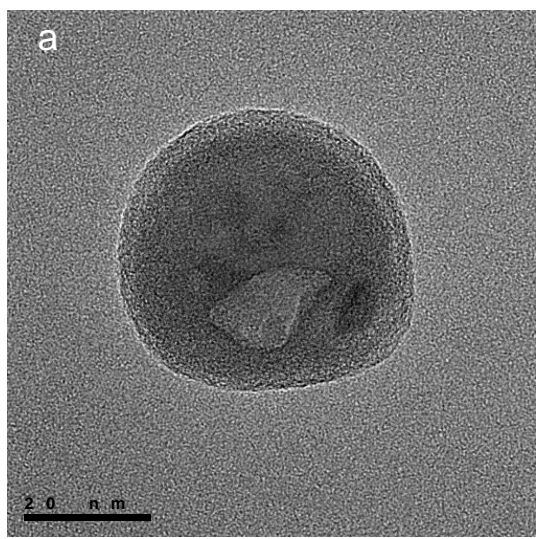

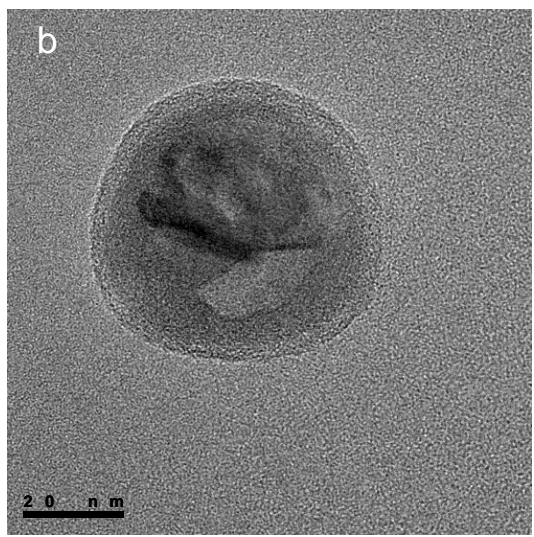

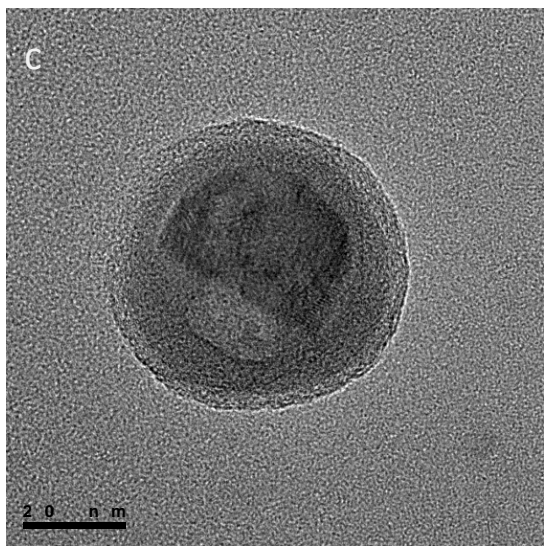

Fig. 2. HR TEM image of an aluminium particle observed with different tilt angles: (a) zero tilt of the grid; (b) tilt of +28.1°; (c) tilt of -29.1°.

High resolution microscopy confirmed that the center of the particle was well crystallized, the (111) and (200) planes corresponding to fcc Al were observed in the same zone by tilting the particle (Fig. 3). These planes have the same orientation for the whole particle, in contrast to our previous results on Al NPs generated in pure ethanol without saturation with $H_2$ [5].

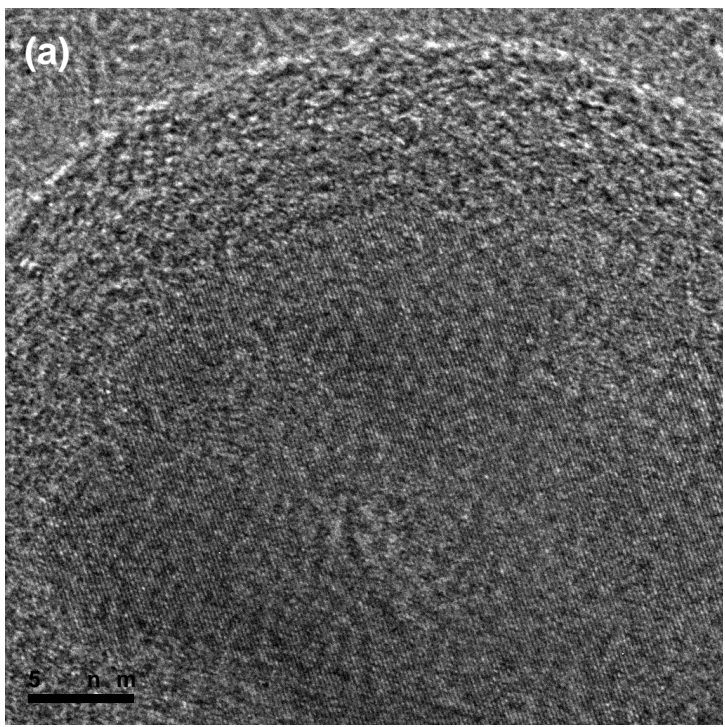
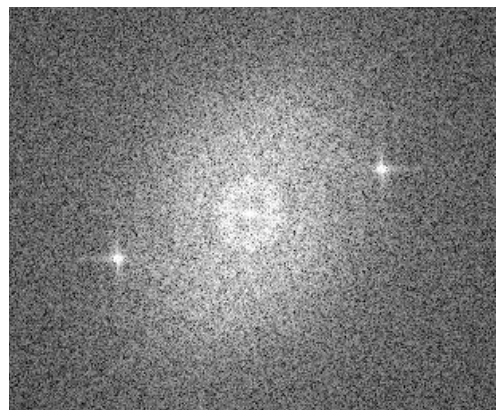
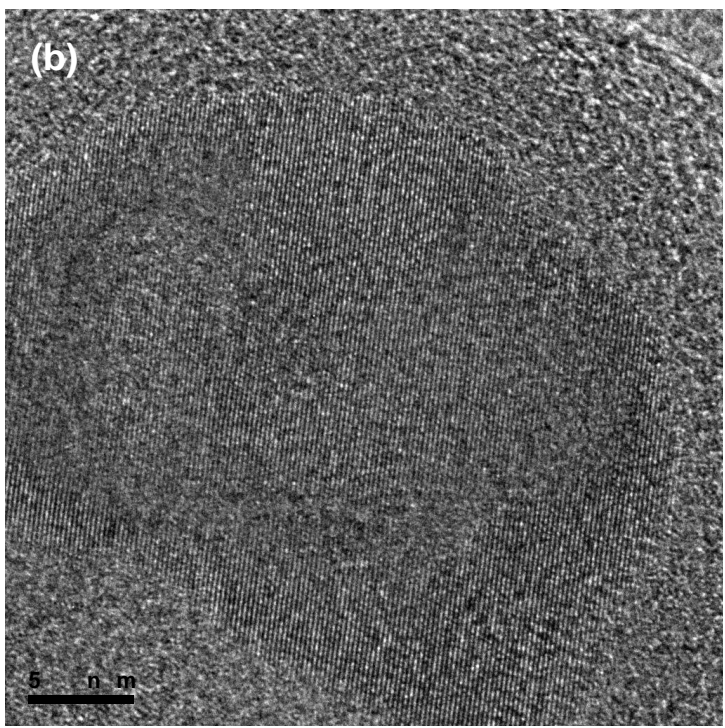
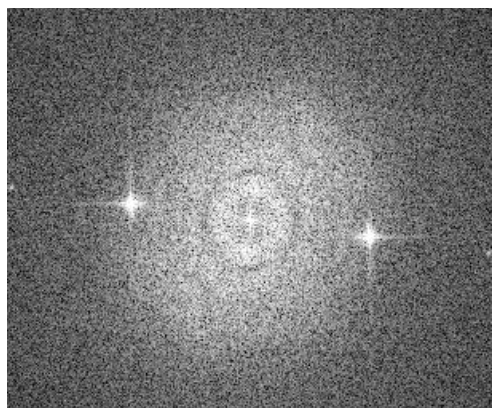

Fig. 3. HR TEM images of an Al particles with different tilt angle : (a) +28,1°; (b) -29,1° . The reticular distances are 2Å (a) and 2.3Å (b) corresponding to the (200) and (111) planes of fcc aluminium Al, respectively. Images to the right are numerical diffraction patterns of corresponding images of NPs.

The core-shell structure is observed more frequently than in our previous work with laser ablation of a bulk Al target in pure ethanol. This structure is summarized in the Figure 4 : the particles are made up of a well crystallized fcc Al inner part (dark zone) embedded with a amorphous metal Al shell (grey zone) and coated by an alumina shell (bright zone). The center of the particle is empty, or, at least has smaller density of the image, tentatively, due to the emission of hydrogen during the solidification of the particle after the end of the laser pulse.

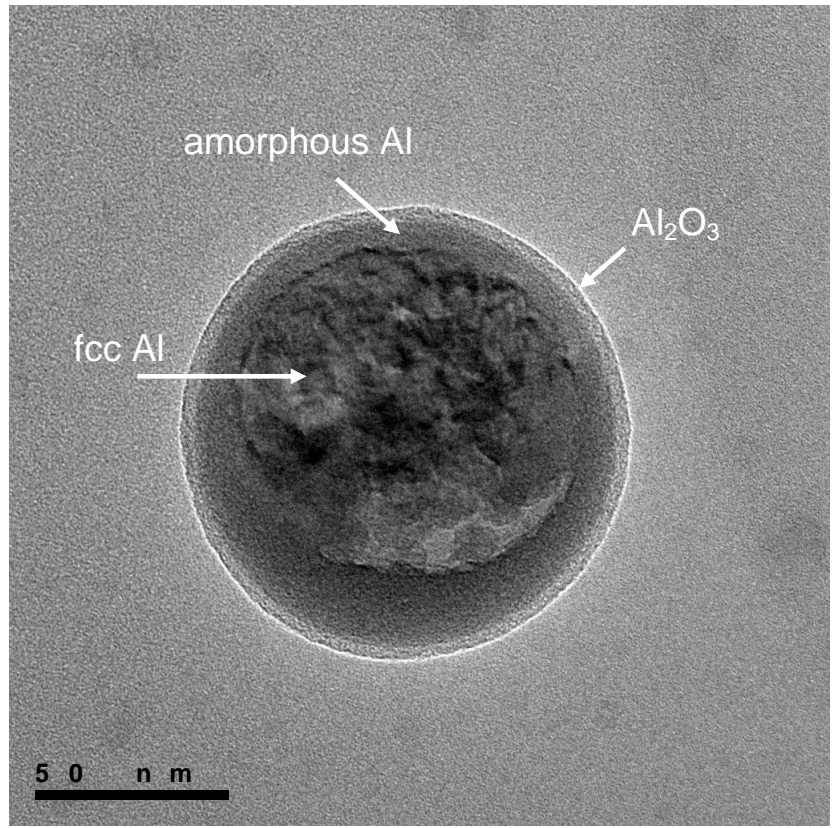

Fig. 4. HR TEM image of a particle showing the metal core, an inner shell of non-crystallized Al and an outer layer of alumina.

This core shell structure can be tentatively related to the mechanism of formation of the NPs. After the ejection from the target, the particle solidifies losing heat through the transfer to surrounding liquid ethanol saturated with hydrogen. The cooling rate of outer parts of the particle is higher than in the middle, so the amorphous Al shell is formed. Solidification of a metal into amorphous state is usually associated with high rate of temperature quenching, of order of $10^{10}$-$10^{12}$ K/s. The cooling rate inside the particle is lower, and the crystalline structure is established. At the same time, $H_2$ dissolved in the particle during its existence in liquid phase is released inside the particle with possible formation of Al hydrides. This release is due to the very sharp dependence of its solubility in Al [6] (see Fig. 5). Molten Al NPs serve as a diffusion sink for the surrounding $H_2$.

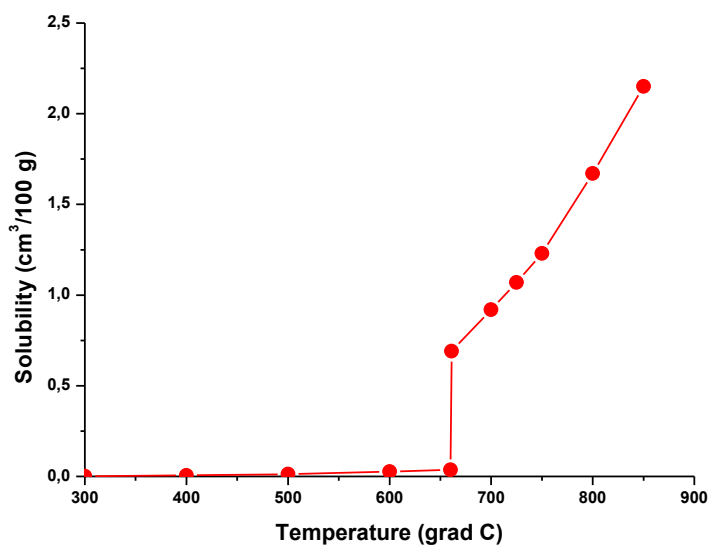

Fig. 5. Temperature dependence of hydrogen solubility in Al (plotted after data in [6]).

The solubility drops by two orders of magnitude upon solidification of Al. The outer layer of oxide in Fig. 4 could appear upon further exposure of the samples to air for the transfer to the TEM chamber. The hydrogen that is released inside the particle upon its solidification provides additional pressure, which induces complete crystallization of the metal inside it. Indeed, unlike our previous observations on laser-synthesized Al NPs, the interior of Al NPs obtained by ablation in ethanol with $H_2$ is a single crystal Al. It seems likely that the cavities in NPs are either filled with gaseous $H_2$ or are made of amorphous material such as Al hydrides. Of course, gaseous $H_2$ may diffuse out from the nanoparticle into surrounding liquid until the equilibrium content in both contacting media is reached. However, the permeability of Al for $H_2$ is extremely low compared to other metals [7], so $H_2$ trapped inside the particle may remain in it for a long time. This point requires further spectroscopic studies, e.g., $^1$H-RMN.

Laser ablation of Al target in liquid ethanol saturated with $H_2$ leads to weak eye-visible reddish coloration of the colloidal solution. The spectrum of the solution shows a wide band at 520 nm (Fig. 6, curve 1). This band may be tentatively assigned to the formation of Al hydrides. This suggestion is partially supported by the observation of the same absorption band under purging $H_2$ through the colloidal solution of Al NPs at 50° C prepared by laser ablation in liquid ethanol without $H_2$ (Fig.6. curve 2). Since gas purging could hardly alter the size distribution of Al NPs, one may conclude that the absorption in the visible observed on both samples is due to formation of Al hydride.

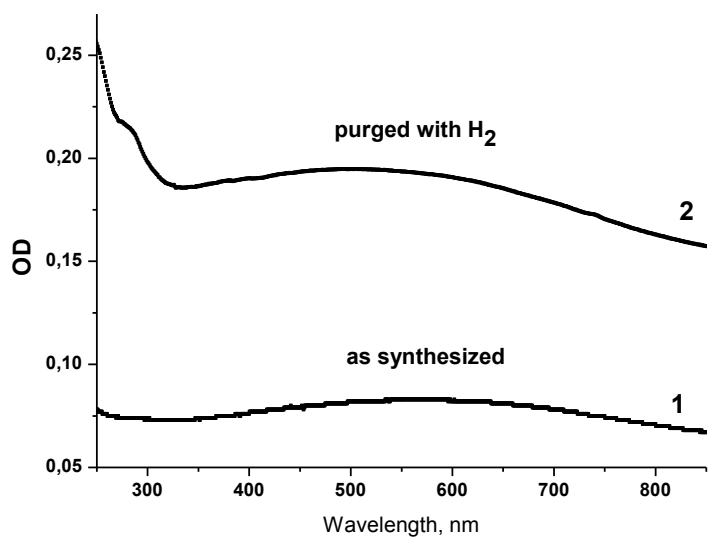

Fig. 6. Optical density of the colloidal solution of Al NPs in ethanol in visible range of spectrum. As synthesized (1) and after purging gaseous $H_2$ during 15 minutes at 50° C (2).

$H_2$ may form around the Al NPs during their synthesis by laser ablation not only due to the reaction of molten Al with traces of water as it was suggested in [5] but also due to partial decomposition of ethanol. Reddish coloration of the colloidal solution of Al NPs in presence of $H_2$ is corroborated with the description of a sample of stabilized $AlH_3$ [8].

In conclusion, hydrogen-rich NPs of metallic Al have been synthesized by laser ablation of a bulk Al target in liquid ethanol saturated with $H_2$ at atmospheric pressure. NPs have a distinct core-shell structure as inferred from HR TEM microscope observation. The core is made of alumina and a layer of amorphous Al. The next shell consists of single crystal Al. Most of NPs have a cavity inside. Its formation is tentatively assigned to the release of hydrogen during the solidification of Al NPs due to very sharp dependence of hydrogen dissolution in Al with temperature.